\documentclass[journal]{IEEEtran}

\usepackage{amsmath,amssymb,amsfonts}
\usepackage{booktabs}
\usepackage{array}
\usepackage{cite}
\usepackage{url}
\usepackage{hyperref}
\usepackage{graphicx} 
\usepackage{float} 

\hypersetup{
    colorlinks=true,
    linkcolor=black,
    citecolor=black,
    urlcolor=black
}

\newcommand{\sech}{\operatorname{sech}}

\title{A Capacity-Aware Parr Model for Agile Projects}

\author{Pedro E. Colla\\
\textit{UADER-FCyT}\\
\href{mailto:colla.pedro@uader.edu.ar}{colla.pedro@uader.edu.ar}%
\thanks{R\&D effort part of the project UADER-PI/B 230/24.}}

\begin{document}

\maketitle

\begin{abstract}
Classical software effort distribution models, including the Putnam--Norden--Rayleigh family and Parr's alternative curve, were designed to describe the time distribution of development effort under an implied staffing pattern. Their direct use in agile environments is limited when team capacity is fixed, partially fixed, or externally constrained: the original curve may prescribe a staff demand that the organization cannot allocate. This paper proposes a compact refactoring of Parr's model as a capacity-aware forecasting layer for agile projects. The contribution is deliberately narrower than a full causal theory of project dynamics. A normalized Parr-shaped latent effort demand is combined with an observed or planned capacity trajectory. The resulting model forecasts aggregate progress, completion time, capacity deficit, and capacity slack without assuming that the same internal activity path is followed under resource restriction. The model uses a small parameter set: total effort $K$, a Parr shape parameter $\alpha$, an origin constant $c$ that can match nonzero initial staffing, and the capacity trajectory $C(t)$. A discrete sprint formulation is provided, together with a calibration method from ordinary Scrum records and a rolling-origin validation protocol against simple management baselines.
\end{abstract}


\section{Introduction}
Software effort estimation has long combined two different questions that are often confused in practice: how much effort a project will require, and how that effort will be distributed over time. Parametric models such as COCOMO address the first problem through cost drivers and size estimates \cite{boehm1981}, while effort distribution models such as the Putnam--Norden--Rayleigh model (PNR) describe aggregate staffing or effort profiles over the project life cycle \cite{putnam1978}. Parr proposed an alternative to the Rayleigh curve by deriving a symmetric effort density from a logistic view of revealed and unrevealed work \cite{parr1980}. Subsequent analysis observed both the mathematical appeal of the curve and the difficulty of using it as a resource estimation instrument without careful calibration and empirical support \cite{basili1981}.

Agile delivery settings make the problem more visible. Scrum teams work in bounded sprints, inspect progress frequently, and select work based on past performance, upcoming capacity, and the \textit{"definition of done"} \cite{schwaber2020}. In such settings, staffing is often not the dependent variable predicted by a curve; rather, capacity is a managerial constraint. A stable team may be assigned to a project before the forecast is produced. A classical Parr curve may therefore prescribe more staff than the organization can allocate, or less staff than the team already has available.

This paper proposes a refactored and deliberately parsimonious model. The aim is not to build a complete causal theory of agile project dynamics. The aim is to make a Parr-shaped effort curve usable as an aggregate forecasting component when capacity is externally constrained. The proposed model treats the Parr curve as a \emph{latent demand of effort}, not as a mandatory staffing policy. Observed or planned capacity controls how fast this latent demand can be consumed. The model therefore supports three management uses: initial duration forecast, sprint-by-sprint reforecast, and scenario analysis for capacity changes.

\section{Background and Positioning}

The PNR model represents software effort through a Rayleigh-shaped profile and has been widely used in macro-level software sizing and estimating \cite{putnam1978}. Parr's model replaces the Rayleigh assumption with a curve derived from a logistic relationship between work already revealed and work remaining to be revealed \cite{parr1980}. If $K$ is total effort, \textit{$\alpha$ is the curve shape parameter}, and $c$ shifts the origin, the classical Parr effort rate is proportional to
\begin{equation}
    \frac{K\alpha}{4}\sech^2\left(\frac{\alpha t+c}{2}\right).
\end{equation}
The parameter $c$ is important because the visible project may start after some latent preparation or discovery has already occurred. It also permits nonzero staff demand at the recorded origin.

Basili and Beane compared the Parr curve with other staffing distributions and emphasized the practical difficulty of estimating the curve and validating its staffing implications \cite{basili1981}. This criticism is central to the present work. Rather than claiming that Parr provides a complete staffing law, this paper uses it as a smooth, low-dimensional latent demand curve that must be calibrated and tested against observed agile data.

\subsection{Agile Estimation Context}
Agile effort estimation remains dominated by expert judgment, story-point practices, historical velocity, and increasingly data-driven methods \cite{cohn2005,fernandezdiego2020,pasuksmit2024}. Recent reviews continue to report estimation challenges related to information quality, team factors, process variation, changing requirements, and insufficient historical data \cite{pasuksmit2024,ruk2025}. Replication studies of automated agile effort estimation also warn that complex models may fail to outperform simple baselines under realistic validation settings \cite{tawosi2023}. This motivates a conservative baseline-oriented validation protocol.

The proposed model is not a replacement for sprint planning, expert estimation, or backlog refinement. It is a project-level forecasting layer that connects total effort, planned capacity, and aggregate progress. It is also not a machine-learning model for estimating individual user stories. Its intended granularity is the project or release trajectory. The Table~\ref{tab:comparison} summarizes the intended comparison. The paper does not attempt to reproduce every alternative effort distribution in detail. Instead, it positions the contribution against representative families that a reviewer would naturally expect.

\begin{table}[t]
\caption{Schematic comparison with related model families.}
\label{tab:comparison}
\centering
\footnotesize
\begin{tabular}{p{0.18\linewidth}p{0.25\linewidth}p{0.25\linewidth}p{0.20\linewidth}}
\toprule
Model family & Main role & Capacity treatment & Role in this paper \\
\midrule
Rayleigh / PNR & Macro effort distribution over time \cite{putnam1978} & Usually implicit in the fitted profile & Baseline distribution family \\
Parr & Logistic-derived effort distribution \cite{parr1980} & Not explicit in the original curve & Latent demand source \\
Gompertz / Richards & Alternative asymmetric growth curves \cite{gompertz1825,richards1959} & Not project-capacity specific & Optional empirical alternatives \\
Velocity baseline & Agile completion forecast from past throughput \cite{cohn2005} & Capacity is embedded in velocity & Main practical baseline \\
Data-driven estimation & Prediction from historical issue or project data \cite{fernandezdiego2020,tawosi2023} & Learned indirectly from data & External benchmark class \\
Proposed model & Capacity-aware aggregate forecast & Explicit capacity trajectory & Main contribution \\
\bottomrule
\end{tabular}
\end{table}

The proposed model should be interpreted as a phenomenological aggregate forecasting mechanism rather than a detailed representation of agile project dynamics. Its use of the Parr curve reflects the practical need for a smooth, low‑dimensional latent demand profile rather than an assumption that the logistic form fully captures discovery, rework, or coordination effects. The model therefore focuses on aggregate progress under externally constrained capacity and does not attempt to represent backlog dependency, architectural coupling, defect generation, or other internal path‑dependent phenomena. This clarification reinforces that the curve shape is selected for parsimony and empirical convenience, not as a universal staffing law.

\section{Use Case and Modeling Assumptions}

The organization might want to forecast aggregate completion time and understand where the available team capacity is likely to constrain progress.

In that regard the model assumes that:
\begin{enumerate}
    \item The total project effort $K$ is available as an initial estimate, a planning value, or a retrospective calibration value.
    \item A low-dimensional Parr-shaped curve is a useful approximation of aggregate latent effort demand.
    \item Capacity is externally observed or planned and may differ from the demand implied by the latent curve.
    \item The state variable is aggregate completion, not a detailed network of tasks, dependencies, or defects.
    \item Forecasting is performed at project or release level, not at individual user-story level.
\end{enumerate}
Based on this set of assumptions an intended use case for this model is to start with an initial estimate of total project effort $K$, a planned or observed capacity trajectory and one or more historical scrum records at sprint granularity and obtain a forecast of the dynamic evolution of the main project variables, specially effort. 
Also, with less expected precision, the model might be applied to forecast the evolution of a project  given a partial history from itself. 

It's likely that when available resources are constrained from the actual resources demand the total duration will increase, it's hard to forecast if that increase will preserve some sort of path invariance or will introduce inefficiencies, therefore the analysis of such behaviour is left as part of a future work and kept outside the scope of the present work.

\section{Capacity-Aware Parr Formulation}

Let $\tau\geq 0$ denote a latent effort-maturity clock. The normalized Parr-shaped effort density is defined as
\begin{equation}
    g_{\alpha,c}(\tau)=
    \frac{\frac{\alpha}{4}\sech^2\left(\frac{\alpha\tau+c}{2}\right)}{A(c)},
    \label{eq:g_density}
\end{equation}
where the factor $A(c)$ is a normalization constant introduced to ensure that the shifted Parr density integrates to one over the project time domain $t \geq 0$. Since the parameter $c$ shifts the curve relative to the project origin, part of the unshifted curve may lie before $t=0$. Therefore,

\begin{equation}
A(c)=
\int_{0}^{\infty}
\frac{\alpha}{4}
\operatorname{sech}^{2}
\left(
\frac{\alpha t+c}{2}
\right)
dt
=
\frac{1-\tanh(c/2)}{2}
\label{eq:normalization}
\end{equation}

Dividing by $A(c)$ makes the latent effort-demand profile a proper normalized density on the observable project interval $[0,\infty)$.

Thus
\begin{equation}
    \int_0^\infty g_{\alpha,c}(\tau)d\tau=1.
\end{equation}
For total effort $K$, the latent effort demand rate is
\begin{equation}
    D_{\alpha,c,K}(\tau)=K g_{\alpha,c}(\tau)
    \label{eq:latent_demand_tau}
\end{equation}

The corresponding cumulative latent progress is
\begin{equation}
    F_{\alpha,c}(\tau)=
    \frac{
    \tanh\left(\frac{\alpha\tau+c}{2}\right)-\tanh(c/2)
    }{1-\tanh(c/2)}
    \label{eq:cumulative}
\end{equation}
This function maps the latent clock into aggregate completion $x\in[0,1]$.

To obtain a state representation let use
\begin{equation}
    x(t)\in[0,1]
\end{equation}
be the fraction of total effort completed at real time $t$. The latent clock associated with the current state is
\begin{equation}
    \tau(x)=F_{\alpha,c}^{-1}(x).
\end{equation}
The demand rate at state $x$ is therefore
\begin{equation}
    D_{\alpha,c,K}(x)=K g_{\alpha,c}(\tau(x)).
    \label{eq:state_demand}
\end{equation}

The origin constant $c$ is retained as a substantive model element. If the recorded project starts with nonzero assigned staff $C_0$, the model can either choose $c$ to match the initial demand approximately,
\begin{equation}
    c^{\star}=\arg\min_c
    \left[D_{\alpha,c,K}(0)-C_0\right]^2,
    \label{eq:c_initial}
\end{equation}
or estimate $c$ jointly with $\alpha$ using a penalty for initial mismatch. This is important in agile projects where the initial team may already exist at $t=0$.

To introduce the management decision on the resources allocated to the project as a capacity restriction, 
let $C(t)$ be the available capacity rate, measured in the same effort unit per time unit as $D_{\alpha,c,K}$. The effective effort completion rate is
\begin{equation}
    R(t)=\min\left\{C(t),D_{\alpha,c,K}(x(t))\right\}.
    \label{eq:effective_rate}
\end{equation}
The aggregate project dynamics are
\begin{equation}
    \frac{dx(t)}{dt}=\frac{R(t)}{K},\qquad x(0)=0.
    \label{eq:state_dynamics}
\end{equation}

Equation~\eqref{eq:effective_rate} has a direct management interpretation. When latent demand exceeds available capacity, the project is capacity-constrained. When available capacity exceeds latent demand, the excess is not automatically transformed into progress. It may represent slack, coordination limits, work not yet ready, or capacity that can be reassigned. This avoids the unrealistic conclusion that any additional staffing necessarily compresses the project.

Two diagnostic quantities follow immediately:
\begin{align}
    \Delta^-(t) &= \left[D_{\alpha,c,K}(x(t))-C(t)\right]_+,\label{eq:deficit}\\
    \Delta^+(t) &= \left[C(t)-D_{\alpha,c,K}(x(t))\right]_+ .\label{eq:slack}
\end{align}
Here $\Delta^-$ is capacity deficit and $\Delta^+$ is capacity slack. These diagnostics are useful even when the final duration forecast is uncertain.

Because the logistic tail is asymptotic, practical completion is defined by a threshold
\begin{equation}
    x(t)\geq q,
    \label{eq:threshold}
\end{equation}
where $q$ may be chosen as $0.95$, $0.98$, or $0.99$ depending on the desired planning convention. This threshold must be reported explicitly in empirical validation.

\section{sprint-Level Formulation}

For Scrum data, the model is used in discrete time. Let $k=0,1,\ldots$ index sprints and let $\Delta t_k$ be sprint duration. Let $C_k$ be the average capacity rate during sprint $k$. Then
\begin{equation}
    x_{k+1}=\min\left\{1,
    x_k+\frac{\Delta t_k}{K}
    \min\left(C_k,D_{\alpha,c,K}(x_k)\right)
    \right\}.
    \label{eq:discrete_model}
\end{equation}
The predicted effort completed during sprint $k$ is
\begin{equation}
    \hat e_k=K(x_{k+1}-x_k).
    \label{eq:predicted_effort}
\end{equation}
The predicted capacity deficit and slack are
\begin{align}
    \hat\Delta^-_k &= \Delta t_k\left[D_{\alpha,c,K}(x_k)-C_k\right]_+,\label{eq:sprint_deficit}\\
    \hat\Delta^+_k &= \Delta t_k\left[C_k-D_{\alpha,c,K}(x_k)\right]_+.\label{eq:sprint_slack}
\end{align}

If capacity is recorded directly as effort available per sprint rather than as effort per time unit, the product $\Delta t_k C_k$ can be replaced by the observed sprint capacity $\bar C_k$ and the discrete equation becomes
\begin{equation}
    x_{k+1}=\min\left\{1,
    x_k+\frac{1}{K}
    \min\left(\bar C_k,\bar D_{\alpha,c,K}(x_k)\right)
    \right\},
    \label{eq:sprint_capacity_model}
\end{equation}
where $\bar D_{\alpha,c,K}$ is the latent demand expressed per sprint. This version might be more convenient for industrial datasets.

Having established the calibration mechanism, we now evaluate the model’s empirical behavior under realistic validation protocols using real project datasets.

\section{Calibration From Scrum Records}

\subsection{Available Data}
The proposed calibration assumes ordinary project records rather than special experimental instrumentation. For project $j$, the minimal dataset is
\begin{equation}
    \mathcal{D}_j=\{k,\Delta t_{jk},C_{jk},e_{jk},sp_{jk}\}_{k=1}^{n_j},
\end{equation}
where $\Delta t_{jk}$ is the sprint time interval, $C_{jk}$ is average staff or effort capacity, $e_{jk}$ is observed effort, and $sp_{jk}$ is completed story points. Total effort is
\begin{equation}
    K_j=\sum_{k=1}^{n_j} e_{jk}
\end{equation}
for retrospective calibration. In true forecasting, $K_j$ must be replaced by an estimate available at the forecast origin.

The preferred observed progress variable is effort-based:
\begin{equation}
    x^{obs}_{jk}=\frac{\sum_{i=1}^{k}e_{ji}}{K_j}.
    \label{eq:observed_progress_effort}
\end{equation}
If effort is unavailable but completed story points are consistently recorded, a story-point proxy may be used:
\begin{equation}
    x^{obs,sp}_{jk}=\frac{\sum_{i=1}^{k}sp_{ji}}{\sum_{i=1}^{n_j}sp_{ji}}.
    \label{eq:observed_progress_sp}
\end{equation}
The two definitions should not be silently mixed; if both are used, they should be reported as separate analyses.

\subsection{Project-Level Parameter Fit}
For a single historical project, $\alpha$ and $c$ are estimated by minimizing the discrepancy between observed and modeled cumulative progress:
\[
\begin{aligned}
e^x_{jk}(\alpha,c)
&=x^{obs}_{jk}
-x^{mod}_{jk}(\alpha,c;K_j,C_{j,1:k}),\\
p_{j0}(\alpha,c)
&=
\frac{D_{\alpha,c,K_j}(0)-C_{j0}}{K_j}.
\end{aligned}
\]
\[
J_j(\alpha,c)=
\sum_{k=1}^{n_j}w_{jk}
\left[e^x_{jk}(\alpha,c)\right]^2
+
\omega_0\left[p_{j0}(\alpha,c)\right]^2.
\]
\begin{equation}
    (\hat\alpha_j,\hat c_j)=
    \arg\min_{\alpha>0,c}
    J_j(\alpha,c).
    \label{eq:project_fit}
\end{equation}
The first term fits the progress trajectory. The second term is optional and enforces consistency with initial assigned staff when $C_{j0}$ is reliable. The weights $w_{jk}$ may be uniform or may emphasize later forecast accuracy.

This objective deliberately uses cumulative progress rather than only instantaneous staff. sprint-level effort and staff observations are noisy; cumulative progress is usually more stable and closer to the project management question of interest.
When several projects are available, the organization can estimate a robust common value
\begin{equation}
    \alpha^{org}=\operatorname{median}\left(\hat\alpha_1,\ldots,\hat\alpha_N\right),
    \label{eq:org_alpha}
\end{equation}
while allowing $c_j$ to vary by project. This reflects the interpretation that $\alpha$ is a rough organizational or domain shape parameter, whereas $c$ captures the recorded origin and initial allocation of a specific project.

A hierarchical estimation approach could be used later, but the first empirical version should prefer robust summaries and confidence intervals over additional latent parameters. This is consistent with the paper's central design constraint that the model should be calibrable from standard historical scrum records.

The model explicitly does \emph{not} assume that the internal project path remains invariant under restriction. It does not claim that dependencies, rework, communication, or architectural choices are unchanged when the team is over- or under-staffed. Those effects are real, but they are outside the main model and should appear as forecast error, recalibration need, or future model extensions.
The model relies on several operational assumptions that should be stated explicitly. First, the Parr curve is used as an approximate latent demand signal; its suitability in agile environments depends on its stability and low parameter count rather than on empirical proof of logistic effort revelation. Second, the conversion from staff to effective capacity is treated as linear for calibration purposes, although real projects may exhibit nonlinear or context dependent productivity. Third, the model is sensitive to errors in the total effort estimate K, which directly affects the scale of the latent demand and the resulting forecast. These sensitivities do not invalidate the formulation but should be acknowledged when interpreting results.

Several complementary approaches exist in the literature, including throughput based forecasting, Little’s Law formulations, and stochastic flow models used in Kanban style environments. These approaches emphasize work in progress dynamics and queueing behavior rather than latent demand curves. Their integration is outside the scope of this paper, but they provide useful context for understanding the limitations of purely curve based aggregate forecasting. The present model is positioned as a compact alternative that explicitly incorporates capacity trajectories while retaining the interpretability of classical effort distribution families.

\subsection{Project baseline fit}

For a single historical Scrum project, the calibration does not estimate a universal organizational law. It estimates the pair of parameters $(\alpha,c)$ that makes the capacity-constrained Parr trajectory reproduce, as closely as possible, the observed cumulative progress of that project.

Let the project contain $n$ sprints. Let $\bar C_k$ denote the effective capacity available during sprint $k$, expressed in effort units per sprint, and let $e_k$ denote the effort actually recorded during that sprint. For retrospective calibration, total effort is

\begin{equation}
K=\sum_{k=1}^{n} e_k .
\end{equation}

The observed cumulative progress after sprint $k$ is

\begin{equation}
x^{obs}*k=
\frac{\sum*{i=1}^{k} e_i}{K},
\qquad k=1,\ldots,n .
\label{eq:observed_progress_single_project}
\end{equation}

For each candidate pair $(\alpha,c)$, the model generates a complete simulated trajectory. The simulated state is initialized as

\begin{equation}
x^{mod}_0(\alpha,c)=0 .
\end{equation}

Then, for $k=0,\ldots,n-1$, the next state is computed recursively as

\[
\begin{aligned}
d_k(\alpha,c)
&=
\bar D_{\alpha,c,K}
\left(x^{mod}_{k}(\alpha,c)\right),\\
r_{k+1}(\alpha,c)
&=
\min\left[
\bar C_{k+1},d_k(\alpha,c)
\right].
\end{aligned}
\]
\begin{equation}
\begin{split}
x^{mod}_{k+1}(\alpha,c)
&=
\min\left\{
1,
 x^{mod}_{k}(\alpha,c)
+
\frac{r_{k+1}(\alpha,c)}{K}
\right\}.
\end{split}
\label{eq:model_progress_recursion}
\end{equation}

The term $\bar D_{\alpha,c,K}(x)$ is the latent Parr demand expressed in effort units per sprint at aggregate state $x$. To make the calibration operational, it is convenient to write this demand directly as a function of $x$. Let

\begin{equation}
b=\tanh(c/2),
\qquad
z(x)=b+(1-b)x .
\end{equation}

Using the normalized Parr cumulative curve, the demand at state $x$ can be written as

\begin{equation}
\bar D_{\alpha,c,K}(x)
=
\frac{K\alpha}{2}
\frac{1-z(x)^2}{1-b}.
\label{eq:state_demand_closed_form}
\end{equation}

The project-level fit is then obtained by minimizing the discrepancy between observed and simulated cumulative progress:

\[
\begin{aligned}
e^x_k(\alpha,c)
&=x^{obs}_k-x^{mod}_k(\alpha,c),\\
p_0(\alpha,c)
&=
\frac{\bar D_{\alpha,c,K}(0)-\bar C_0}{K}.
\end{aligned}
\]
\[
J(\alpha,c)=
\sum_{k=1}^{n}w_k
\left[e^x_k(\alpha,c)\right]^2
+
\omega_0\left[p_0(\alpha,c)\right]^2.
\]
\begin{equation}
    (\hat\alpha,\hat c)
    =
    \arg\min_{\alpha>0,c}
    J(\alpha,c).
    \label{eq:project_level_fit}
\end{equation}

The first term fits the cumulative progress trajectory. The second term is optional and penalizes inconsistency between the initial model demand and the initial observed capacity. Setting $\omega_0=0$ gives a pure trajectory fit; using $\omega_0>0$ is appropriate when the initial staff allocation is considered reliable and should be reflected by the origin parameter $c$.

This calibration should be interpreted carefully. With a single project, $\hat\alpha$ and $\hat c$ are project-level fitted parameters. They are useful for explaining the observed trajectory and for initializing later organizational calibration, but they should not yet be interpreted as stable organizational constants. A robust organizational value of $\alpha$ requires several projects or a rolling-origin validation design.

\subsection{Rolling Reforecast}
For use during execution, the model is updated at sprint $r$. The observed state is
\begin{equation}
    x_r^{obs}=\frac{\sum_{i=1}^{r}e_i}{\hat K_r},
\end{equation}
where $\hat K_r$ is the total effort estimate available at sprint $r$. The forecast then simulates \eqref{eq:discrete_model} from $r$ forward using a future capacity scenario. Two modes should be kept separate:
\begin{itemize}
    \item \emph{Ex-post capacity isolation}: actual future capacity is used only to test the shape model, not as a deployable forecast.
    \item \emph{True forecast}: future capacity is estimated from information available at sprint $r$, for example planned allocation or recent average capacity.
\end{itemize}
This distinction prevents data leakage and makes the validation credible.

\section{Validation Protocol}

The empirical validation should answer whether the capacity-aware Parr model improves useful management forecasts compared with simpler alternatives. The unit of validation is the project, not a randomly selected sprint. Random sprint-level splits would mix training and future information from the same project and would overstate performance.

\subsection{Baselines}
The minimum baselines are:
\begin{enumerate}
    \item \emph{Constant velocity}: remaining story points divided by historical average velocity.
    \item \emph{Constant capacity}: remaining effort divided by historical average capacity.
    \item \emph{Unconstrained Parr}: the same normalized Parr curve without the capacity minimum in \eqref{eq:effective_rate}.
    \item \emph{Naive median duration}: historical median remaining duration from comparable projects, if enough projects exist.
\end{enumerate}
A Rayleigh/PNR curve can also be included as a shape baseline when sufficient data are available.

\subsection{Forecast Metrics}
For completion-time prediction, the primary metrics are
\begin{align}
    MAE_T &= \frac{1}{M}\sum_{m=1}^{M}\left|\hat T_m-T_m\right|,\label{eq:mae}\\
    RMSE_T &= \sqrt{\frac{1}{M}\sum_{m=1}^{M}\left(\hat T_m-T_m\right)^2}.\label{eq:rmse}
\end{align}
For progress trajectory prediction,
\begin{equation}
    RMSE_x=\sqrt{\frac{1}{\sum_j n_j}\sum_j\sum_k\left(x^{obs}_{jk}-x^{mod}_{jk}\right)^2}.
    \label{eq:rmse_progress}
\end{equation}
If probabilistic forecasts are generated by bootstrap resampling or parameter sampling, interval coverage should be reported:
\begin{equation}
    Coverage_p=\Pr\left(T_{real}\in[\hat T_{p/2},\hat T_{1-p/2}]\right).
\end{equation}

When a very limited history information is available or even if only one scrum project dataset is available , the model cannot be externally validated as an organizational forecasting law. However, the dataset can still be used for three limited but useful purposes: project-level calibration, operational verification, and rolling-origin internal validation.

Let the project contain $n$ sprints. For each sprint $k$, the dataset contains the effective capacity $\bar C_k$, the recorded effort $e_k$, and optionally the completed story points $sp_k$. For retrospective project-level calibration, the total effort is computed as

\begin{equation}
K=\sum_{k=1}^{n}e_k .
\end{equation}

The observed cumulative effort progress after sprint $k$ is

\begin{equation}
x^{obs}*k=
\frac{\sum*{i=1}^{k}e_i}{K},
\qquad k=1,\ldots,n .
\label{eq:single_project_observed_progress}
\end{equation}

For each candidate parameter pair $(\alpha,c)$, the model is run over the same sprint capacity sequence observed in the project. The modeled progress starts at

\begin{equation}
x^{mod}_0(\alpha,c)=0 .
\end{equation}

Then, for $k=0,\ldots,n-1$,

\[
\begin{aligned}
d^{s}_k(\alpha,c)
&=
\bar D_{\alpha,c,K}
\left(x^{mod}_{k}(\alpha,c)\right),\\
r^{s}_{k+1}(\alpha,c)
&=
\min\left[
\bar C_{k+1},d^{s}_k(\alpha,c)
\right].
\end{aligned}
\]
\begin{equation}
\begin{split}
x^{mod}_{k+1}(\alpha,c)
&=
\min\left\{
1,
 x^{mod}_{k}(\alpha,c)
+
\frac{r^{s}_{k+1}(\alpha,c)}{K}
\right\}.
\end{split}
\label{eq:single_project_model_progress}
\end{equation}

The fitted project-level parameters are obtained by minimizing the discrepancy between observed and modeled cumulative progress:

\[
\begin{aligned}
e^x_k(\alpha,c)
&=x^{obs}_k-x^{mod}_k(\alpha,c),\\
p_0(\alpha,c)
&=
\frac{\bar D_{\alpha,c,K}(0)-\bar C_0}{K}.
\end{aligned}
\]
\[
J(\alpha,c)=
\sum_{k=1}^{n}w_k
\left[e^x_k(\alpha,c)\right]^2
+
\omega_0\left[p_0(\alpha,c)\right]^2.
\]
\begin{equation}
    (\hat\alpha,\hat c)
    =
    \arg\min_{\alpha>0,c}
    J(\alpha,c).
    \label{eq:single_project_fit}
\end{equation}

The first term fits the cumulative trajectory. The optional second term penalizes inconsistency between the initial latent demand and the initial observed capacity. This is useful when the initial staff allocation is considered a reliable signal of the project state at the beginning of the recorded Scrum trajectory.

After calibration, operational correctness is assessed by checking that the fitted model satisfies the expected structural properties: cumulative progress must be monotone, bounded by one, limited by available capacity, and able to reproduce the observed completion range within a specified tolerance. These checks do not validate generalization, but they verify that the capacity-constrained mechanism behaves consistently on the available project trace.

A stronger internal assessment can be performed through rolling-origin validation. For a sequence of forecast origins $r<n$, only the data available up to sprint $r$ are used to estimate or update the model state. The model then forecasts the remaining trajectory from sprint $r+1$ onward. For each origin $r$, the forecasted completion sprint is

\begin{equation}
\hat T^{(r)}
=
\min
\left\{
k \geq r :
x^{mod}_{k\mid r} \geq q
\right\},
\label{eq:rolling_forecast_completion}
\end{equation}

where $q$ is a practical completion threshold, such as $q=0.99$. The forecast error at origin $r$ is

\begin{equation}
\varepsilon_T^{(r)}
=
\hat T^{(r)} - T^{obs},
\label{eq:rolling_forecast_error}
\end{equation}

where $T^{obs}$ is the observed completion sprint. Across all forecast origins, the rolling-origin errors can be summarized using the metrics defined above, such as $MAE_T$, $RMSE_T$, and $RMSE_x$.

With a single project, this procedure should be interpreted as internal case-study validation. It evaluates whether the model can be calibrated from an ordinary Scrum trace and whether it produces coherent reforecasts as more sprint information becomes available. It does not establish that the fitted parameters are stable across projects. External validation requires multiple independent project datasets.

\subsection{Single-Project Temporal Calibration Protocol}

The numerical validation reported in this paper follows a temporal holdout design using a single real historical scrum project, the project comprised the delivery of 1033 story points involving 5424 hours of effort across 22 sprints of 2 weeks uniform duration each. The staff involved was defined by management varying between 5 and 8
persons per sprint, the staff allocation for each sprint was fairly constant within the sprint. On first
analysis the information about two of the sprints revealed strong deviations due to special causes, therefore
that information was removed from the analysis.  

The available dataset contains, for each sprint, the sprint identifier, delivered Story Points, recorded effort, and average assigned staff. Since only one project is available, the experiment should be interpreted as an internal case-study validation rather than as external organizational validation.

Let the project contain $n$ sprints, indexed by $k=1,\ldots,n$. The sprint identifier is treated as an integer time index. The dataset is first sorted by sprint number, and the first $r$ sprints are used for calibration. Unless otherwise stated, $r=\lfloor n/2 \rfloor$. The remaining sprints, $r+1,\ldots,n$, are used only for validation.

The procedure is applied twice. In the first run, the response variable is recorded effort. In the second run, delivered story Points are used as a proxy for project progress. Let

\begin{equation}
y_k \in {e_k, sp_k}
\end{equation}

denote the selected response variable, where $e_k$ is the effort recorded in sprint $k$ and $sp_k$ is the number of Story Points delivered in that sprint.

For retrospective validation of the model mechanism, total project size is computed as

\begin{equation}
K_y = \sum_{k=1}^{n} y_k .
\label{eq:retrospective_total_size}
\end{equation}

In a purely operational forecast, $K_y$ should instead be supplied as an external planned estimate. The retrospective value in \eqref{eq:retrospective_total_size} is used here only to normalize progress and to evaluate the behavior of the model under a known project total.

The observed cumulative progress after sprint $k$ is

\begin{equation}
x^{obs}*k =
\frac{\sum*{i=1}^{k} y_i}{K_y}.
\label{eq:observed_cumulative_progress_validation}
\end{equation}

A key normalization step is the conversion of staff into effective capacity. Since staff is not expressed in the same units as effort or Story Points, a capacity conversion rate is estimated using only the calibration segment:

\begin{equation}
\rho_y =
\frac{\sum_{k=1}^{r} y_k}
{\sum_{k=1}^{r} Staff_k}.
\label{eq:capacity_conversion_rate}
\end{equation}

The planned capacity for sprint $k$ is then

\begin{equation}
\bar C^{(y)}_k = \rho_y Staff_k .
\label{eq:capacity_from_staff}
\end{equation}

This assumption means that future effort or future Story Points are not used to construct the forecast. In the validation segment, only the future staff allocation is used, through \eqref{eq:capacity_from_staff}, as the planned capacity signal.

For each candidate pair $(\alpha,c)$, the constrained Parr model is simulated over the calibration segment. The modeled state is initialized as

\begin{equation}
x^{mod}_0(\alpha,c)=0.
\end{equation}

For $k=1,\ldots,r$, the model computes the sprint-level modeled delivery as

\begin{equation}
\hat y_k(\alpha,c)
=
\min\Bigl(
\bar C^{(y)}*k,
\bar D*{\alpha,c,K_y}
\bigl(x^{mod}_{k-1}(\alpha,c)\bigr)
\Bigr),
\label{eq:calibration_delivery_rule}
\end{equation}

and updates cumulative progress according to

\begin{equation}
x^{mod}*k(\alpha,c)
=
x^{mod}*{k-1}(\alpha,c)
+
\frac{\hat y_k(\alpha,c)}{K_y}.
\label{eq:calibration_state_update}
\end{equation}

The fitted project-level parameters are obtained by minimizing a combined discrepancy between observed and modeled progress over the calibration segment:

\[
\begin{aligned}
e^x_k(\alpha,c)
&=x^{obs}_k-x^{mod}_k(\alpha,c),\\
e^y_k(\alpha,c)
&=
\frac{y_k-\hat y_k(\alpha,c)}{K_y},\\
p^{(y)}_0(\alpha,c)
&=
\frac{\bar D_{\alpha,c,K_y}(0)-\bar C^{(y)}_1}{K_y}.
\end{aligned}
\]
\[
\begin{aligned}
J_T(\alpha,c)
&=
\sum_{k=1}^{r}w_x
\left[e^x_k(\alpha,c)\right]^2
+
\sum_{k=1}^{r}w_y
\left[e^y_k(\alpha,c)\right]^2\\
&\quad+
\omega_0\left[p^{(y)}_0(\alpha,c)\right]^2
+
\eta(\alpha-\alpha_0)^2.
\end{aligned}
\]
\begin{equation}
    (\hat\alpha,\hat c)
    =
    \arg\min_{\alpha>0,c}
    J_T(\alpha,c).
    \label{eq:temporal_project_fit}
\end{equation}

The first term fits the cumulative progress trajectory. The second term fits sprint-level delivery. The third term is an optional initial-capacity penalty that encourages consistency between the initial latent demand and the initial observed staff allocation. The final term is a weak shape regularization used to avoid extreme values of $\alpha$ when the latent demand is only partially identifiable under binding capacity constraints.

After calibration, the fitted parameters $(\hat\alpha,\hat c)$ are kept fixed. The model is then evaluated over the validation segment. In this segment, the forecast is generated from the planned capacity clock induced by future staff:

\begin{equation}
\hat y_k =
\min\Bigl(
\bar C^{(y)}*k,
K_y-\hat Y*{k-1}
\Bigr),
\qquad k=r+1,\ldots,n,
\label{eq:validation_capacity_clock}
\end{equation}

where

\begin{equation}
\hat Y_k = \hat Y_{k-1}+\hat y_k
\end{equation}

is the modeled cumulative delivered work. The latent Parr demand remains available as a diagnostic estimate of required capacity, but future delivery during validation is driven only by the staff-based capacity plan. This separation prevents the validation forecast from using future effort or future Story Points while still allowing the model to evaluate whether the planned staff allocation is sufficient.

Errors are computed separately for the calibration and validation segments. For sprint-level prediction, the validation RMSE is

\begin{equation}
RMSE_{val}
=
\sqrt{
\frac{1}{n-r}
\sum_{k=r+1}^{n}
\left(
y_k-\hat y_k
\right)^2
}.
\label{eq:validation_rmse}
\end{equation}

To express the magnitude of the error in relative terms, the root mean squared percentage error is also computed:

\begin{equation}
RMSPE_{val}
=
\sqrt{
\frac{1}{n-r}
\sum_{k=r+1}^{n}
\left(
100
\frac{\hat y_k-y_k}{y_k}
\right)^2
}.
\label{eq:validation_rmspe}
\end{equation}

sprints with zero observed delivery require either exclusion from percentage-error metrics or a small denominator regularization. In the present dataset, the percentage metrics are computed only when the observed value is positive.

The graphical protocol follows the same temporal separation. Original project data are displayed as blue points. The fitted constrained model over the calibration segment is displayed as a solid red line. The validation forecast is displayed as a dotted red line. The optional unconstrained Parr baseline, when shown, is displayed as a solid green line. The calibration cutoff sprint is marked explicitly, and sprint numbers are treated as integer time points.

The first half of the sprints data sequence were used to obtain a calibration of the model whilst the second half is used 
to compare the actual results with the forecast made by the model and evaluate the deviation shown. As the total project size and capacity allocation is considered an external management decision the values of the 
provided dataset are used.

The results obtained from the validation can be seen in the following figures both for the forecast of effort
and story points delivered, also the RMS deviation as a percentage is shown for the effort, the deviation for story points is similar and consistent in terms of error with it.

\begin{figure}[H]
    \centering
    \includegraphics[width=0.95\linewidth]{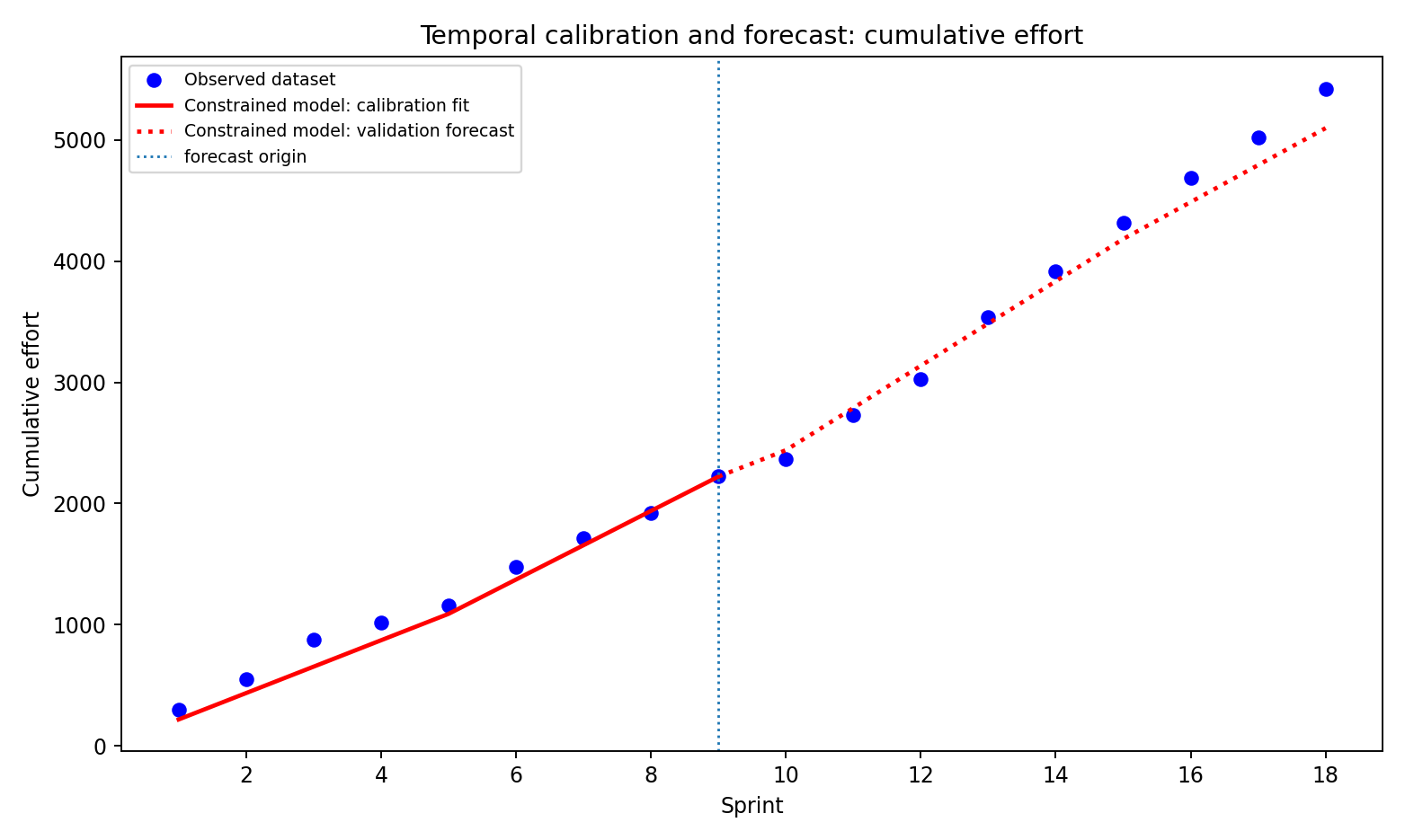}
    \caption{Comparison effort forecast vs. actual}
    \label{fig:run_001a}
\end{figure}


\begin{figure}[H]
    \centering
    \includegraphics[width=0.95\linewidth]{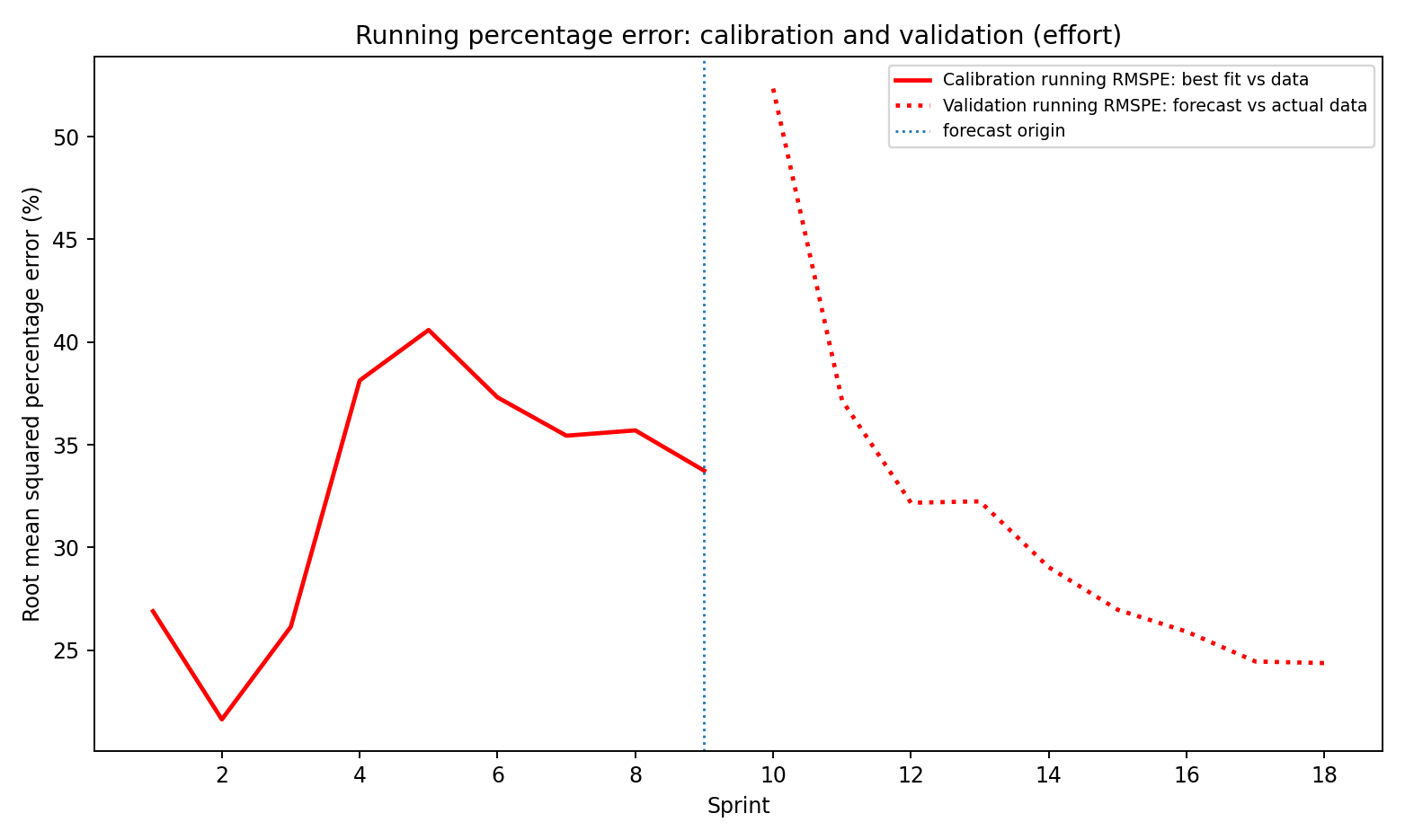}
    \caption{Evaluated RMS as a percent between actual and forecasted effort}
    \label{fig:run_001c}
\end{figure}

Even if the above figures refers to results using effort as the predicted outcome of the model also story 
points can be used with that purpose with similar results.

\subsection{Monte Carlo Validation}

To complement the model temporal behaviour validation, a Monte Carlo emulation is used to evaluate the sensitivity of the model results to variations on the project management decisions and the influence of the project context in the model calibrated parameters.

A first run is executed where the model is calibrated with the entire set of project values and the resulting
configuration is used for a sensitivity analysis to understand the impact in variations of the different factors, the result can be seen in the following tornado graphic figure.

\begin{figure}[H]
    \centering
    \includegraphics[width=0.95\linewidth]{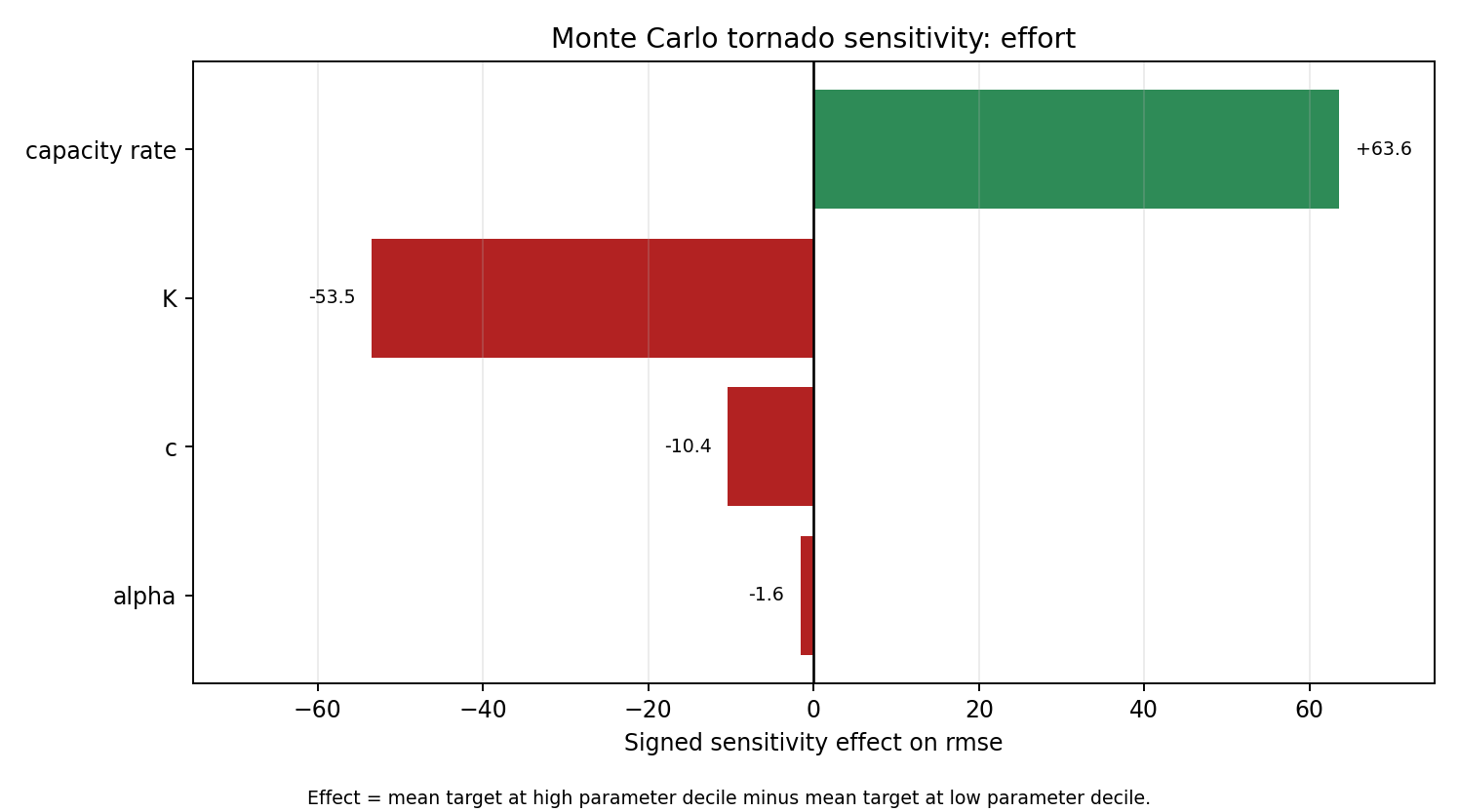}
    \caption{Sensitivity of the forecasted effort to variations in the model parameters}
    \label{fig:run_002a}
\end{figure}

The main influence in the outcome, somewhat to be expected, is on variations of the project scope and the capacity assigned by management being changes on the calibration factors less significant.

Capacity uncertainty was modeled separately. Instead of sampling capacity from a continuous interval, the staff signal was sampled from the discrete set of staff values actually observed in the project dataset.

A second evaluation uses the already used calibration approach to use half the project data to calibrate the
model and the other half as part of the model being turned an oracle of the project outcome. In this run distorsions are made in the model parameters using an uniform distribution for it's values and also 
an stochastic assignment of capacity within the limits of the values shown by the project reference data, 
the intent is to understand likely variances observed in the outcome as well as special sensitivity to
a random variation of the context.

The result can be seen 

\begin{figure}[H]
    \centering
    \includegraphics[width=0.95\linewidth]{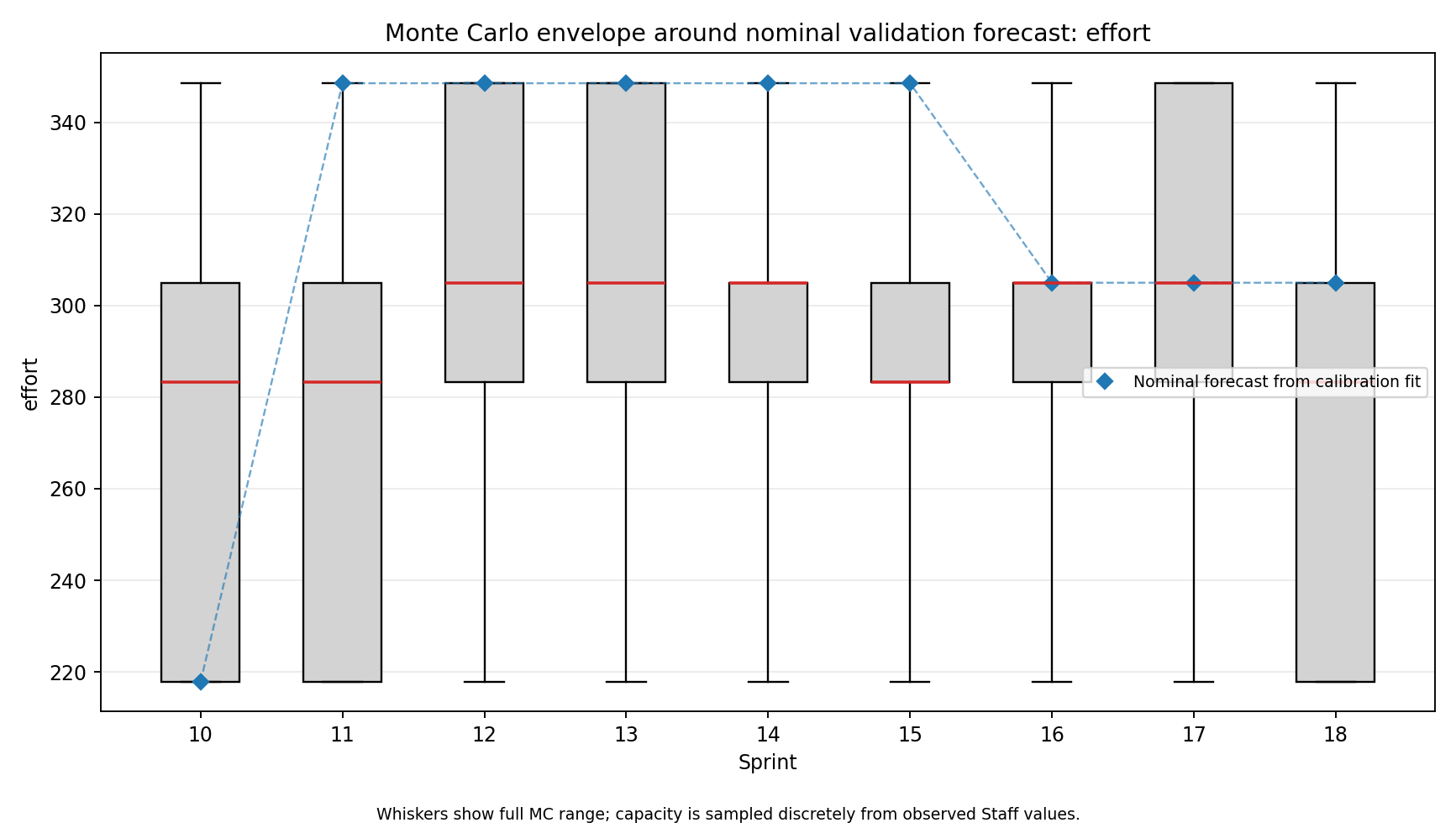}
    \caption{Sensitivity of the forecasted effort to the capacity and variations in the model parameters}
    \label{fig:run_002b}
\end{figure}

The resulting distribution of $\hat y^{MC}_{j,k}$ was summarized with boxplots for each validation sprint. The nominal forecast $\hat y^{nom}_k$, obtained from the calibrated model using the actual capacity signal of the dataset, was overlaid as the reference point. Therefore, the graph does not compare the Monte Carlo distribution against the real project observations; instead, it evaluates how stable the model forecast is under stochastic perturbations of its parameters and capacity signal.

This validation is diagnostic rather than predictive in the strict statistical sense. Its purpose is to determine whether the calibrated forecast is robust to plausible variation in the model parameters and in the capacity signal. When the nominal forecast lies inside the Monte Carlo range, the calibrated model behavior is consistent with the uncertainty envelope induced by the selected perturbation assumptions.

Overall, the validation demonstrates internal coherence and operational plausibility, while highlighting the need for multi‑project datasets for external generalization.

\subsection{Success Criteria}
The model should not be accepted merely because it fits historical curves. A useful empirical result would require at least one of the following:
\begin{enumerate}
    \item lower rolling-origin completion-time error than constant velocity and constant capacity baselines;
    \item better progress trajectory error than unconstrained Parr when capacity is visibly restrictive;
    \item diagnostically useful identification of capacity deficit or slack periods confirmed by project records;
    \item calibrated uncertainty intervals with reasonable empirical coverage.
\end{enumerate}
The validation should also report negative results. If the model does not outperform simple baselines, that is evidence that the added structure is not justified for the dataset.

The empirical validation presented in this paper is intentionally limited. Because only a single historical Scrum project is available, the results should be interpreted as internal case study validation rather than external organizational generalization. The aim is to demonstrate calibrability, internal coherence, and operational plausibility of the capacity aware mechanism. The model is not claimed to outperform velocity based or capacity based baselines in general; instead, it provides a structured way to connect total effort, capacity plans, and aggregate progress. Future validation with multiple independent projects is required to assess parameter stability, comparative accuracy, and organizational applicability.

Overall, the validation demonstrates internal coherence and operational plausibility, while highlighting the need for multi‑project datasets for external generalization.

\section{Discussion}

The core advantage of the proposed formulation is that it retains the useful part of Parr---a compact effort-distribution shape---while changing its managerial interpretation. The curve no longer prescribes how many people must be assigned at each instant. Instead, it describes a latent aggregate demand that may or may not be matched by available capacity. This makes the model compatible with agile teams whose staffing is stable, negotiated, or constrained by organizational reality.

The second advantage is empirical discipline. Previous attempts to extend effort curves can easily accumulate parameters for overload, rework, dependency deformation, and organizational inefficiency. Such mechanisms are plausible, but they are difficult to calibrate from ordinary Scrum records. The present formulation intentionally keeps these effects outside the main model. They may be studied later as residual patterns, not inserted upfront as unidentifiable parameters.

The third advantage is managerial interpretability. Forecasts are expressed in terms of expected completion, capacity deficit, and capacity slack. These quantities can support decisions such as preserving a stable team, adding temporary capacity, reducing scope, delaying commitments, or reallocating idle capacity. Even when the completion forecast is uncertain, the deficit and slack diagnostics may be actionable.

Assumptions around the theoretical evolution of the effort curve has been kept to a minimum for several reasons. First, the model avoids the strong assumption that a resource-constrained project follows the same internal dependency path as an unconstrained project. Second, it keeps the main formulation small enough to be calibrated from ordinary historical scrum records such as sprint, story points, effort, and average staff or capacity. Third, it separates the forecasting model from optional explanations of rework, coordination loss, or architectural dependency. Finally, it defines a validation protocol based on rolling-origin forecasts rather than retrospective curve fitting alone.

\section{Limitations and Threats to Validity}

The model has several limitations that should be stated explicitly. It does not model deformation of the latent demand curve under severe capacity restriction, even though such deformation may occur in practice. It assumes independence between latent demand and available capacity, omitting potential feedback loops such as delayed discovery or increased rework. It also depends on the quality of effort and capacity records, and on the accuracy of the total effort estimate K. These limitations reflect the deliberately narrow scope of the contribution and motivate future extensions.

The model is aggregate. It does not represent backlog dependency, architectural coupling, defect dynamics, rework loops, team learning, or coordination loss. System-dynamics models address some of these effects at the cost of many additional assumptions and parameters \cite{abdelhamid1991,madachy2008}. The present model deliberately does not compete with such models as a causal simulation of project behavior.

The model also depends on the quality of $K$. If total effort is badly estimated, the forecast will inherit that error. In retrospective calibration, using the final observed effort as $K$ is acceptable for estimating $\alpha$ and $c$, but not for claiming true predictive performance. Rolling-origin validation must use only the total-effort estimate available at the forecast date.

Story points introduce additional threats. They are team-specific, may drift over time, and are not necessarily comparable across projects. Therefore, effort-based progress should be preferred whenever effort records are available. Story-point progress can be useful, but only under consistent local estimation practice.

Finally, capacity is not equivalent to productivity. Two teams with the same headcount may differ in skill, availability, domain knowledge, and interruption load. The model treats capacity as the effective effort rate available to the project. If the organization records only nominal staff, an additional preprocessing step is required to translate staff into effective capacity.

\section{Conclusion}

This paper refactors the extension of Parr's effort model for agile use by narrowing the claim and simplifying the mathematics. The proposed model combines a normalized Parr-shaped latent effort demand with an explicit capacity trajectory. It forecasts aggregate progress and completion time while diagnosing capacity deficit and slack. The formulation preserves the origin constant $c$, allowing the model to represent nonzero initial staffing, but avoids the stronger and less defensible assumption that a restricted project follows the same internal activity path as an unrestricted project.

Future research should include multi project validation, comparative evaluation against velocity and Rayleigh/PNR baselines, and sensitivity analysis of parameters $\alpha$ and c. Extensions may incorporate stochastic capacity, Bayesian estimation of K, or deformable curve families that adapt to capacity constraints. Integrating rework dynamics or backlog flow models may also improve realism while preserving the aggregate nature of the formulation.

Among the activities also left as future work are to enhance validation with additional data used with empirical execution to implement the sprint-level calibration, evaluate rolling-origin forecasts on historical Scrum datasets, and compare the results against constant velocity, constant capacity, unconstrained Parr, and Rayleigh/PNR baselines. The model should be retained only if it improves forecast usefulness or provides capacity diagnostics that are not available from simpler baselines.

\end{document}